\documentclass{article}
\usepackage{ijcai21}

\usepackage{times}
\usepackage{soul}
\usepackage{url}
\usepackage[hidelinks]{hyperref}
\usepackage[utf8]{inputenc}
\usepackage[small]{caption}
\usepackage{graphicx}
\usepackage{amsthm}
\usepackage{booktabs}
\usepackage{algorithm}
\usepackage{algorithmic}

\usepackage{amsmath,amssymb,amsfonts}
\usepackage{textcomp}
\usepackage{xcolor}

\usepackage{epsfig}
\usepackage{bbm}
\usepackage{multirow}

\usepackage{subcaption}
\title{Detection of Adversarial Supports in Few-shot Classifiers Using Self-Similarity and Filtering}

\author{

Yi Xiang Marcus Tan$^{1,2}$
\and
Penny Chong$^{1,2}$\and
Jiamei Sun$^{2}$\and
Ngai-Man Cheung$^{1,2}$\and
Yuval Elovici$^{1,3}$\And
Alexander Binder$^{4}$
\affiliations

$^1$ST Engineering-SUTD Cyber Security Laboratory\\
$^2$ISTD pillar, Singapore University of Technology and Design, Singapore\\
$^3$Department of Software and Information Systems Engineering, Ben-Gurion University, Israel\\
$^4$ Department of Informatics, University of Oslo, Norway
\emails
\{marcustanyx16,pennychong94,sunjiamei.hit\}@gmail.com, 
ngaiman\_cheung@sutd.edu.sg,\\
elovici@bgu.ac.il, 
alexabin@ifi.uio.no
}

\newif\ifmark
\newcommand{\changemarker}[1]{%
\ifmark
\textcolor{blue}{#1}%
\else
#1%
\fi
}

\usepackage[normalem]{ulem}

\begin{document}
\markfalse 

\maketitle

\begin{abstract}
Few-shot classifiers excel under limited training samples, making them useful in applications with sparsely user-provided labels. Their unique relative prediction setup offers opportunities for novel attacks, such as targeting support sets required to categorise unseen test samples, which are not available in other machine learning setups. In this work, we propose a detection strategy to identify adversarial support sets, aimed at destroying the understanding of a few-shot classifier for a certain class. We achieve this by introducing the concept of self-similarity of a support set and by employing filtering of supports. Our method is attack-agnostic, and we are the first to explore adversarial detection for support sets of few-shot classifiers to the best of our knowledge. Our evaluation of the miniImagenet (MI) and CUB datasets exhibits good attack detection performance despite conceptual simplicity, showing high AUROC scores. We show that self-similarity and filtering for adversarial detection can be paired with other filtering functions, constituting a generalisable concept.
\end{abstract}

\section{Introduction}

An open topic in machine learning is the transferability of a trained model to a new set of prediction categories without retraining efforts, in particular when some classes have very few samples.
Few-shot learning algorithms have been proposed to address this, where prediction and training are based on the concept of an episode. 
Unlike other setups, the prediction in few-shot models is relative to the support set classes of an episode \cite{MAML:finn2017model,SNAIL:mishra2018a,MTL:sun2019meta}. 
The label categories vary in each episode and training is performed by drawing randomised sets of classes, thus iterating over varying prediction tasks when learning model parameters.
Effectively, this learns a class-agnostic similarity metric which generalises to novel categories \cite{RN:sung2018learning,CAN:hou2019cross}.


Unfortunately, the adversarial susceptibility of models under the few-shot paradigm remains relatively unexplored, albeit gaining traction \cite{xu2020meta,goldblum2019robust}. This is compared to models under the standard classification setting, where such phenomenon had been widely explored \cite{szegedy2013intriguing,madry2018towards}.
The relative nature of predictions in few-shot setups allows going beyond crafting adversarial test samples.

The attacker could craft adversarial perturbations for all $n$-shot support samples of the attacked class and insert them into the deployment phase of the model. The goal is to misclassify test samples of the attacked class regardless of the samples drawn in the other classes. In this work, we consider the impact on the few-shot accuracy of the attacked class, in the presence of adversarial perturbations, even when different samples were drawn for the non-attacked classes. This is a highly realistic scenario as the victim could unknowingly draw such adversarial support sets during the evaluation phase once they were inserted by the attacker. The use of adversarial samples to attack other settings than the one trained for are known as transferability attacks.

Prior methods proposed to mitigate such adverse effects through the lenses of detection \cite{xu2017feature,cintasdetecting} and model robustness \cite{folz2020adversarial,zhang2020interpreting}. Though these methods work well for neural networks under the conventional classification setting,
they will fail on few-shot classifiers due to limited data. 
Furthermore, these defences were not trained to transfer its pre-existing knowledge towards a novel distribution of class samples, contrary to few-shot classifiers.
With the aforementioned drawbacks in mind, we propose a conceptually simple method for performing attack-agnostic detection of adversarial support samples in this setting. 
We exploit the concept of support and query sets of few-shot classifiers to measure the similarity of samples within a support set after filtering. 
We perform this by
randomly splitting the original support set randomly into auxiliary support and query sets, followed by filtering the auxiliary support and predicting on the query.
If the samples are not self-similar, we will flag the support set as adversarial.
To this end, we make the following contributions in our work:
\begin{enumerate}
    \item We propose a novel attack-agnostic detection mechanism against adversarial support sets in the domain of few-shot classification. This is based on self-similarity under randomised splitting of the support set and filtering, and is the first, to the best of our knowledge, for the detection of adversarial support sets in few-shot classifiers. 
    \item We investigate the effects of a unique white-box adversary against few-shot frameworks, through the lens of transferability attacks. Rather than crafting adversarial query samples similar to standard machine learning setups, we optimise adversarial supports sets, in a setting where all non-target classes are varying.
    \item We provide further analysis on the detection performance of our algorithm when using differing filtering functions and also different formulation variants of the aforementioned self-similarity quantity.
\end{enumerate}

The remaining of our paper is structured as follows: Section~\ref{related} discusses prior literature and Section~\ref{background} provides readers with the background to this study. We dive into our method in Section~\ref{method} and describe our experimental settings and evaluation results in Section~\ref{exps}. We provide further in-depth discussion in Section~\ref{discuss} and we conclude in Section~\ref{conclude} with summary and future work.

\section{Related Works}
\label{related}


\textbf{Poisoning of Support Sets}: There is limited literature examining the poisoning of support sets in meta-learning. \cite{xu2020meta} proposed an attack routine, Meta-Attack, extending from the highly explored Projected Gradient Descent (PGD) attack \cite{madry2018towards}. They assumed a scenario where the attacker is unable to obtain feedback from the classification of the query set. Hence, the authors used the empirical loss on the support set to generate adversarial support samples to induce misclassification behaviours to unseen queries. 

\subsection{Autoencoder-based and Feature Preserving-based Defences}
\cite{cintasdetecting} performs detection of such attacks using Non-parametric Scan Statistics (NPSS), based on hidden node activations from an autoencoder.
\cite{folz2020adversarial} proposed using an autoencoder to reconstruct input samples such that only the necessary signals remain for classification. Their method requires fine-tuning the decoder based on the classification loss of the input with respect to the ground truth. However, under the few-shot setting, such fine-tuning based on the classification loss should be avoided as we would require large enough samples from each class for this step.
\cite{zhang2020interpreting} attempts to stabilise sensitive neurons which might be more prone to the effects of adversarial perturbations, by enforcing similar behaviours of these neurons between clean and adversarial inputs.
Their method requires adversarial samples during the training process which potentially makes defending against novel attacks challenging. Hence, we proposed a detection approach that does not make use of any adversarial samples. Though we employed the concept of feature preserving as one of our various filtering functions, our approach is different from \cite{zhang2020interpreting} as it does not suffer from this limitation.
Hence, in our work, we adopted an approach that does not require any labelled data.

\section{Background}
\label{background}

\subsection{Few-shot classifiers Used}

A majority of the 
few-shot classifiers are trained with \emph{episodes} sampled from the training set. Each episode consists of a support set $S=\{x_s, y_s\}_{s=1}^{K*N}$ with $N$ labelled samples per $K$ classes, and a query set $Q=\{x_q\}_{q=1}^{N_q}$ with $N_q$ unlabelled samples from the same $K$ classes to be classified, denoted as a $K$-way $N$-shot task. The metric-based classifiers learn a distance metric that compares the features of support samples $x_s$ and query sample $x_q$ and generates similarity scores for classification. During inference, the episodes are sampled from the test set that has no overlapping categories with the training set.

\begin{figure}
    \centering
    \includegraphics[trim={6cm 98.4cm 13cm 5.3cm},clip,width = 0.45\textwidth]{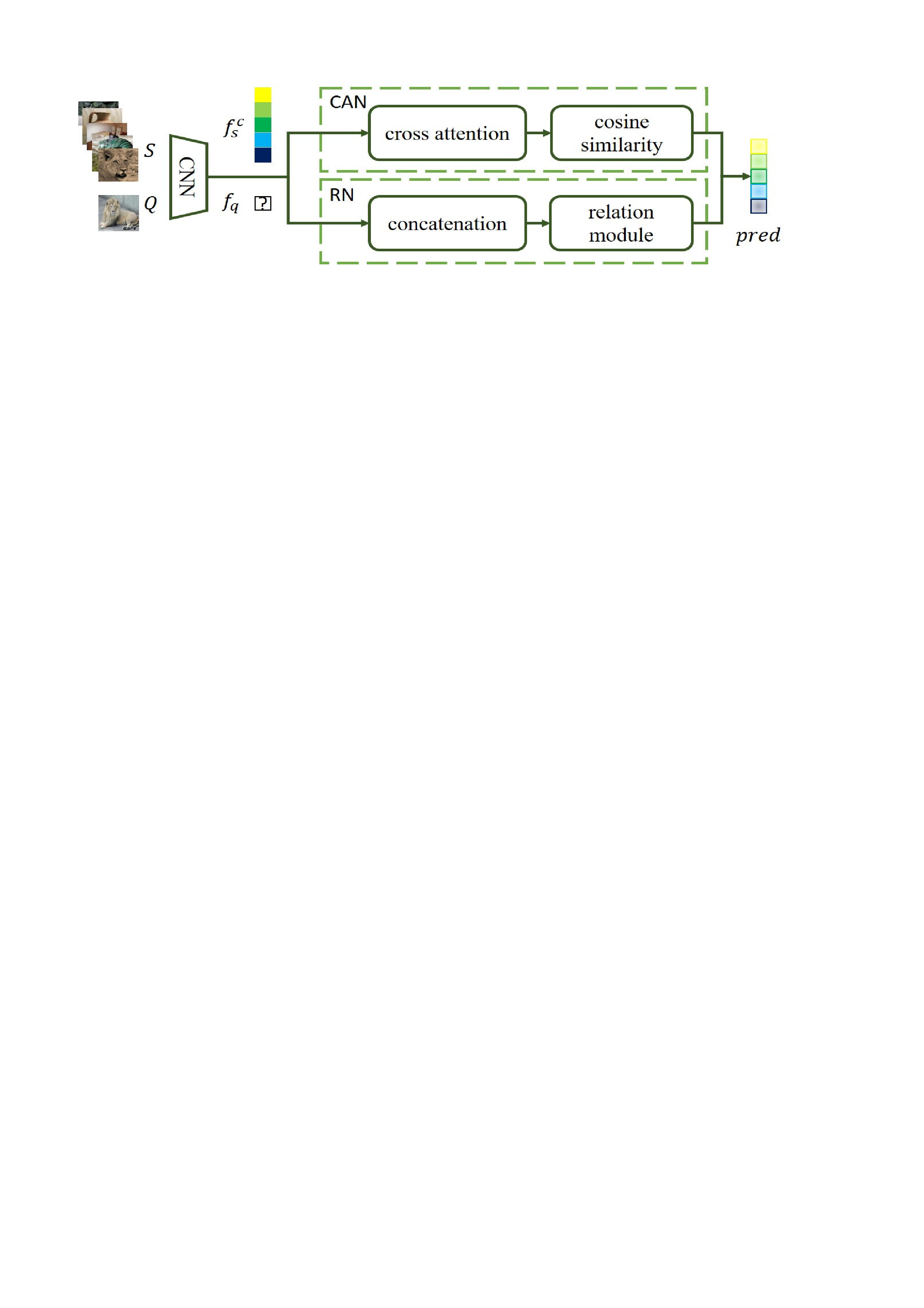}
    \caption{RelationNet and CAN few-shot classifiers.}
    \label{fig:few_shot_classifiers}
\end{figure}

In this work, we explored two known metric-based few-shot classifiers, namely the RelationNet (RN) \cite{RN:sung2018learning} and a state-of-the-art model, the cross-attention network (CAN) \cite{CAN:hou2019cross}. As illustrated in Figure~\ref{fig:few_shot_classifiers}, the support and query samples are first encoded by a backbone CNN to get the image features \{$f^c_s \ | \ c=1,  \dots , K$\} and $f_q$, respectively. The feature vectors $f_s^c$ and $f_q \in \mathbbm{R}^{d_f, h_f, w_f}$, where $d_f$, $h_f$, and $w_f$ are the channel dimension, height, and width of the image features. If $N>1$, $f^c_s$ will be the averaged feature of the support samples from class $c$.
To measure the similarity between $f_s^c$ and $f_q$, the RN model concatenates $f_q$ and $f_s^c$ along the channel dimension pairwise and uses a \emph{relation module} to calculate the similarities. The CAN model adopts a \emph{cross-attention} module that generates attention weights for every $\{f_s^c, f_q\}$ pair. The attended image features are further classified with cosine similarity in the spirit of dense classification \cite{DENSECLASSIFICATION:lifchitz2019dense}.




\subsection{Adversarial Attacks}
Here, we describe the base adversarial attacks used in our experiments.
The PGD attack \cite{madry2018towards} applies the sign of the gradient of the loss function to the input data as adversarial perturbations. It initialises an adversarial candidate by a small noise injection. This process repeats for a number of iterations. For an input $x_i$ at the $i^{th}$ iteration:
\begin{gather}
    x_0 = x_{original} + Uniform(-\epsilon, \epsilon), \\
    x_i = Clip_{x, \epsilon}\{x_{i-1} + \eta \, sign(\nabla_{x} L(h(x_{i-1}), y_{truth}))\},
    \label{pgd}
\end{gather}
where $h(.)$ is a prediction logits for classifier $h$ of some input sample, $y_{truth}$ is the ground truth label, $L$ is the loss used during training (i.e. cross entropy with softmax), $\eta$ is the step size and $\epsilon$ is the adversarial strength which limits the adversarial candidate $x_i$ within an $\epsilon$-bounded $\ell_\infty$ ball.

The Carlini-Wagner (CW) attack \cite{carlini2017towards} finds the smallest $\delta$ that successfully fools a target model using the Adam optimiser. Their attack solves the following objective function:
\begin{equation}
    \begin{split}
    &\min_\delta ||\delta||_2 + const\cdot L(x+\delta, \kappa),\\
    s.t.~L(x', \kappa) &= \max(-\kappa, \max_i(h(x')_{i \neq t}) - h(x')_t).
    \end{split}
    \label{cweqn}
\end{equation}
The first term penalises $\delta$ from being too large while the second term ensures misclassification. The value $const$ is a weighting factor that controls the trade-off between finding a low $\delta$ and having a successful misclassification. 
$h(\cdot)_i$ refers to the logits of prediction index $i$ and $t$ refers to the target prediction. $\kappa$ is the confidence value that influences the logits score differences between the target prediction $t$ and the next best prediction $i$.

\subsection{Threat Model}

We assume that the attacker wants to destroy the few-shot classifier's notion of a targeted class, $t$, unlike conventional machine learning frameworks where one is optimising single test samples to be misclassified. The attacker wants to find an adversarially perturbed set of support images, such that misclassification of most query samples from class $t$ occurs, regardless of the class labels of the other samples. He or she then replaces the defender support set for class $t$ with the adversarial support. We assume that the attacker has white-box access to the few-shot model (i.e. weights, architecture, support set).
The adversarial support set would classify itself as self-similar, that is, they classify among each other as being within the same class, visually appear as class $t$, but classify true query images of class $t$ as belonging to another class.

We now clarify our definition of $x$ used in our attacks. The attacks are applied on a fixed support set candidate $(x_1^{t}, \ldots, x_{n_{shot}}^t)$ for the target class. In every iteration of the gradient-based optimisation, we sample all classes randomly except for the target class.
Specifically, we sample the support sets $S^{-t}$ and query sets $Q^{-t}$ of all the other classes randomly, and we randomly sample the query samples of the target $Q^t$, illustrated in the equations below. They are redrawn in every iteration of the optimisation of the above equations.
\begin{equation}
\begin{split}
 \mathcal{C}^{-t} &\sim Uniform(\mathcal{C}~\backslash~\{t\}) \\
     S^{-t}, Q^{-t} &\sim Uniform(x | c \in \mathcal{C}^{-t}), \ Q^t \sim Uniform(x | c = t)\\
         x &=(x_1^{t}, \ldots, x_{n_{shot}}^t), \\
    h(x)&=   h(x_1^{t}, \ldots, x_{n_{shot}}^t, S^{-t}, Q^t, Q^{-t}),
    \end{split}
    \label{threatmodel}
\end{equation}
where $\mathcal{C}$ is the set of all classes, and $\mathcal{C}^{-t}$ the random set of classes used in the episode together with class $t$. The last line in \eqref{threatmodel} indicates that the few-shot classifier $h$ takes in a support set made up of $x$ and $S^{-t}$ and a query set made up of $Q^t$ and $Q^{-t}$, which is a simplification to the expression, to relate to \eqref{pgd} and \eqref{cweqn}.
The adversarial perturbation $\delta$ and the underlying gradients are computed only for each of the support samples $x$ of the target class.

\section{Defence Methodology}
\label{method}

The defence is based on three components: sampling of auxiliary query and support sets, filtering the auxiliary support sets, and measuring the accuracy on the unfiltered auxiliary query set. We denote a statistic either averaged over all possible splits or for a randomly drawn split of a support set into auxiliary sets with filtering of the auxiliary supports as self-similarity. We elaborate further on auxiliary sets below.


\subsection{Auxiliary Sets}
\label{auxset}
Few-shot classifiers' support and query sets can be freely chosen, implying that any sample can be used as either a support or query. 
Given a support set for class $c$, we randomly split it into auxiliary sets, where $S^c$ might be clean or adversarial:
\begin{gather}
    \nonumber S^{c}_{aux} \cup Q^{c}_{aux} = S^c~and~S^{c}_{aux} \cap Q^{c}_{aux} = \emptyset,\\
    s.t.~|S^{c}_{aux}| = n_{shot}-1~and~|Q^{c}_{aux}| = 1.
\end{gather}
The few-shot learner is now faced with a randomly drawn ($n-1$)-shot problem, evaluating on one query sample per way, with the option to average the $n$ possible splits.






\subsection{Detection of Adversarial Support Sets}
\label{detectsec}

\begin{figure}[htb]
    \centering
    \includegraphics[trim={0cm 0cm 0cm 0cm},clip,width = 0.45\textwidth]{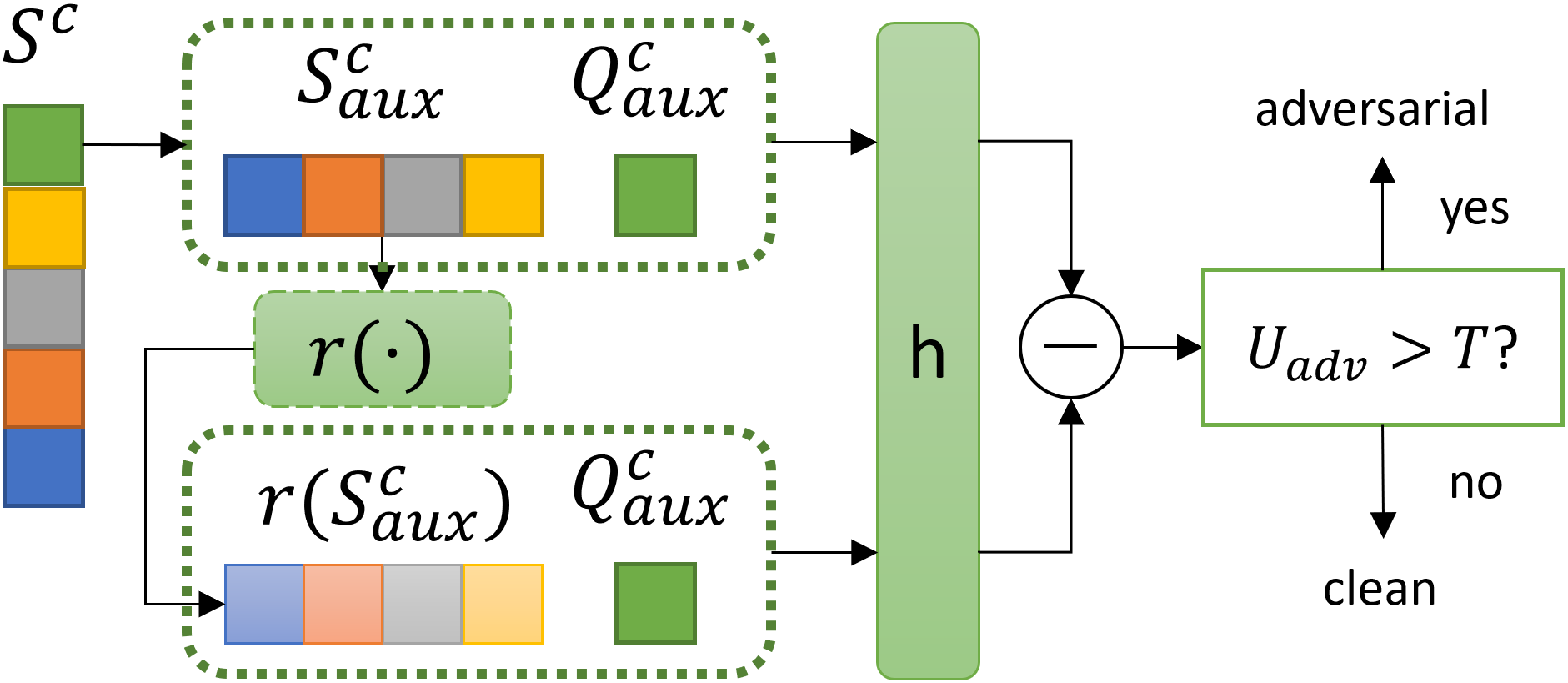}
    \caption{Illustration of our detection mechanism based on self-similarity, by partitioning $S^c$ into two auxiliary sets $S^{c}_{aux}$ and $Q^{c}_{aux}$ and filtering. Best viewed in colour.}
    \label{fig:detectionl1}
\end{figure}

Our detection mechanism flags a support set as adversarial when support samples within a class are highly different from each other, as shown in  Figure~\ref{fig:detectionl1}.
Given a support set of class $c$, $S^c$, we split it randomly into two auxiliary sets $S^{c}_{aux}$ and $Q^{c}_{aux}$. 
We filter $S^{c}_{aux}$ using a function $r(\cdot)$ and use the resultant samples as the new auxiliary support set to evaluate $Q^{c}_{aux}$. 
Following which, we obtain the logits of $Q^{c}_{aux}$ both before and after the filtering of the auxiliary support (i.e. using $S^{c}_{aux}$ and $r(S^{c}_{aux})$ respectively) and compute the $\ell_1$ norm difference between them. 
The adversarial score $U_{adv}$ is given in Eq.~\eqref{ssrate} where $h$ is the few-shot classifier
\begin{equation}
\begin{split}
U_{adv} =&\| h(r(S^c_{aux}),Q^c_{aux} ) - h(S^c_{aux},Q^c_{aux} ) \|_1 ,
\label{ssrate}
\end{split}
\end{equation}
and $r$ is any filtering function which maps a support set onto its own space. 
\changemarker{The filter $r$ is chosen such that it causes smaller impact to clean samples, while inducing larger dissimilarity between the auxiliary support and query sets under adversariality.}
We observe that we obtain already very high AUROC detection scores when computing $U_{adv}$ without averaging over $n_{shot}$ draws, which we elaborate further later. We flag a support set $S_c$ as adversarial if the adversarial score goes above a certain threshold\footnote{\changemarker{$T$ can be chosen by examining the $U_{adv}$ on clean support samples, according to a desired threshold based on False Positive Rates (e.g. @5\% FPR).}} (i.e.~$U_{adv}>T$).
Different statistics can be used to compute $U_{adv}$, with Eq.~\eqref{ssrate} being one of many. Our main contribution lies rather in the proposal of using self-similarity of a support set for such detection.

\subsection{Feature-space Preserving Autoencoder (FPA) for Auxiliary Support Set Filtering}
\label{autoencoder}
We explored using an autoencoder (AE) as a filtering function $r(\cdot)$, for the detection of adversarial samples in the support set, motivated by \cite{folz2020adversarial}. Initially we trained a standard autoencoder to reconstruct the clean samples in the image space using the MSE loss. However, the standard autoencoder performed poorly in detecting adversarial supports since it did not learn to preserve the feature space representation of image samples. Therefore, we switched to a feature-space preserving autoencoder 
which additionally reconstructs the images in the feature space of the few-shot classifier, contrary to prior work where they fine-tuned their AE on the classification loss. We argue that using classification loss for fine-tuning is inapplicable in few shot learning due to having very few labelled samples. 
We minimise the following objective function for the feature-space preserving autoencoder: 
\begin{equation}
\mathcal{L_{FPA}} =\frac{1}{N'}\sum^{N'}_{i=1} 0.01 \cdot \ \frac{\| x_i - \hat{x_i} \|_{2}^{2}}{dim(x_i)^{1/2}} \ + \frac{\| f_i - \hat{f_i} \|_{2}^{2}}{dim(f_i)^{1/2}}
\label{aeeqn}
\end{equation}
where $x_i$ and $\hat{x_i}$ are the original and reconstructed image samples, respectively, and, $f_i$ and $\hat{f_i}$ are the feature representation of the original and reconstructed image obtained from the few-shot model before any metric module (i.e. features from CNN backbone). The second loss term ensures that the reconstructed image features are similar to those of original image in the feature space of the few-shot models. We train the feature-space preserving autoencoder by fine-tuning the weights from the standard autoencoder.

\subsection{Median Filtering (FeatS)}
\label{feats}
In our work, we also explored an alternative filtering function. We adopted a feature squeezing (FeatS) filter from \cite{xu2017feature} where it was used in a conventional classifier. It essentially performs local spatial smoothing of images by having the centre pixel taking the median value among its neighbours within a 2x2 sliding window. As their detection performance was reasonably high using this filter, we decided to use it as an alternative to FPA as an explorative step. However, their approach performs filtering on each individual test sample whereas we use it on the auxiliary support set. Since we would like to demonstrate our detection principle and the FPA performs already very well, we leave further filtering functions to future research.

\section{Experiments and Results}
\label{exps}

\subsection{Experimental Settings}

\textbf{Datasets}: MiniImagenet (MI) \cite{miniImageNet:vinyals2016matching} and CUB \cite{CUB:wah2011caltech} datasets were used in our experiments. We prepared them following prior benchmark splits \cite{LSTMoptimizer:ravi2016optimization,FeaturewiseTranslayer:tseng2020cross}, with 64/16/20 categories for the train/val/test sets of MI and 100/50/50 categories for the train/val/test sets of CUB. In our attack and detection evaluation, we chose an exemplary set of 10 and 25 classes from the test set for MI and CUB respectively, and we report the average metrics across them. This is purely for computational efficiency. For the RN model, we used image sizes of 224 while using image sizes of 96 for the CAN model across both datasets. We shrank the image size for the CAN model due to memory usage issues.

\textbf{Attacks:} In our work, we used two different attack routines, one being PGD while the other being a slight variant of the CW attack. This variant uses a normal Stochastic Gradient Descent optimiser instead of Adam as we did not yield good performing adversarial samples with the latter. We still used the objective function defined in Eq.~\eqref{cweqn} to optimise our CW adversarial samples, while using Eq.~\eqref{pgd} to perform a perturbation step less the clipping and sign functions. We name this attack CW-SGD. For our PGD attack, we limit the $\ell_\infty$ norm of the perturbation to $12/255$ and a step size of $\eta=0.05$ (see Eq.~\eqref{pgd}).
For our CW-SGD attack, clipping was not used due to the optimisation over $||\delta||_2$ while $\kappa=0.1$ and $\eta=50$.
We would like to stress that optimising for the best set of hyperparameters for generating attacks is not the main focus of our work as we are more interested in obtaining viable adversarial samples. In both settings, we generate 50 sets of adversarial perturbations for each of the 10 and 25 exemplary classes for MI and CUB respectively. We also attack all $n$ support samples for the targeted class $t$. 


\textbf{Autoencoder}: We used a ResNet-50 \cite{he2016deep} architecture for the autoencoders\footnote{Autoencoder architecture adapted from GitHub repository https://github.com/Alvinhech/resnet-autoencoder.}. For the MI dataset, we trained the standard autoencoder from scratch with a learning rate of 1e-4. For CUB, we trained the standard encoder initialised from ImageNet with a learning rate of 1e-4, and the standard decoder from scratch with a learning rate of 1e-3. For fine-tuning of the feature-space preserving autoencoder, we used a learning rate of 1e-4.
We employed a decaying learning rate with a step size of 10 epochs and $\gamma=0.1$. We used the Adam \cite{kingma2014adam} optimiser with a weight decay of 1e-4. In both settings, we used the train split for training and the validation split for selecting our best performing set of autoencoder weights out of 150 epochs of training. It is implemented in PyTorch \cite{paszke2017automatic}. 

\subsection{Baseline Accuracy of Few-shot Classifiers}

We evaluated our classifiers by taking the average and standard deviation accuracy over 2000 episodes across all models and datasets, reported in Table~\ref{clf:baseline}, to show that we were attacking reasonably performing few-shot classifiers. 

\begin{table}[htb]
\centering

\begin{tabular}{lcc}
             \hline
             & RN - 5 shot & CAN - 5 shot \\ \hline
MI & 0.727  $\pm$ 0.0037     & 0.787  $\pm$ 0.0033      \\ \hline
CUB          & 0.842  $\pm$ 0.0032     & 0.890  $\pm$ 0.0026     \\ \hline
\end{tabular}
\caption{Baseline classification accuracy of the chosen models on the two datasets, under a 5-way 5-shot setting, computed across 2000 randomly sampled episodes. We report the mean with 95\% confidence intervals for the accuracy.
}
\label{clf:baseline}
\end{table}

\subsection{Attack Evaluation Metrics}

We evaluated the success of our attacks via computing the Attack Success Rate (ASR), measuring the proportion of samples that had adversarial candidates generated from attacks that successfully cause misclassification. We only considered samples from the targeted class when measuring ASR:
\begin{align}
    ASR=\mathbb{E}_{S^t,Q^t \sim D}\{P(\mathrm{argmax}_j(h_j(S^t+\delta^t, Q^t)) \neq t)\}.
\end{align}
The remaining $(K-1)$ classes were sampled randomly. In the evaluation of the detection performances, we used the Area Under the Receiver Operating Characteristic (AUROC) metric, since detection problems are binary (whether an adversarial sample is present or not), and true and false positives can be collected at various predefined threshold values ($T$).

\subsection{Transferability Attack Results}

We conducted transferability experiments to evaluate how well the attacker generalised their generated adversarial perturbation under two unique scenarios: \textit{i) transfer with fixed supports} and \textit{ii) transfer with new supports}.
Setting (i) assumes that we have the same adversarial support set for class $t$ and we evaluated the ASR over newly drawn query sets. Setting (ii) relaxes this assumption, and we instead applied the generated adversarial perturbation, that was stored during the attack phase, on newly drawn support sets for class $t$, similarly evaluating over newly drawn query sets.
Contrary to transferability attacks in conventional setups where a sample is generated on one model and evaluated on another, we performed transferability to new tasks, by drawing randomly sets of non-target classes together with their support sets, and new query sets for the few-shot paradigm.

\begin{figure}
    \centering
    \includegraphics[trim={5.2cm 52cm 9.3cm 7.6cm},clip,width=0.47\textwidth]{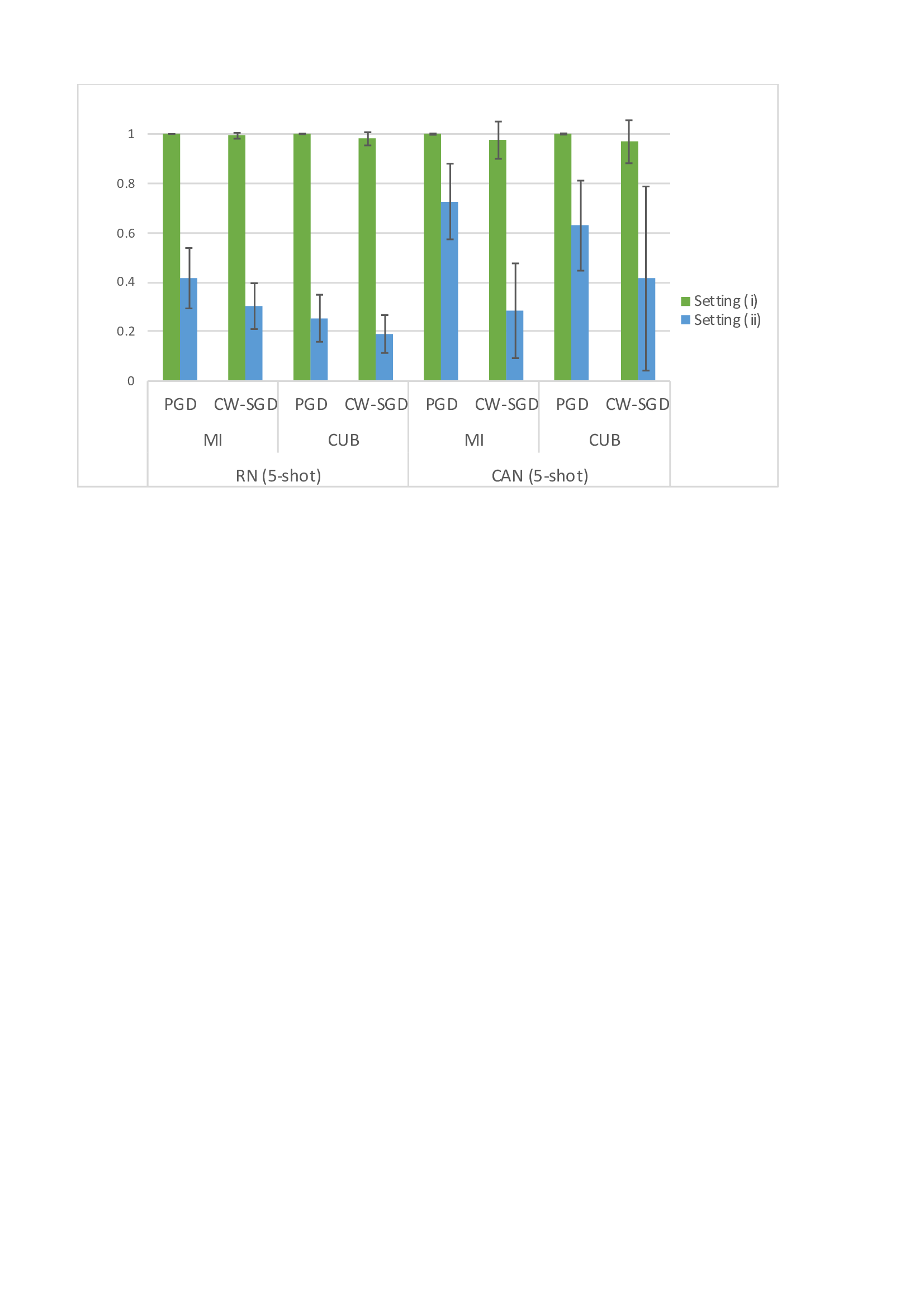}
    \caption{Transferability attack results under scenarios (i) Fixed Supports and (ii) New Supports, against our two explored attacks on RN and CAN models across both datasets. Reported ASRs were averaged across the chosen exemplary classes and across the 50 generated sets of adversarial perturbations as bar charts. Standard deviation represented as the whiskers. ASR metric reported.}
    \label{fig:transfer}
\end{figure}

As illustrated in Figure~\ref{fig:transfer}, the PGD generated adversarial samples showed higher transferability than the CW-SGD attack, across both models and under both scenarios. The exceptionally high transfer ASR we observed under scenario (i) implies that once the attacker had obtained an adversarial support set targeting a specific class, successful attacks can be carried out on new tasks for which the target class is present. This further reinforces the motivation to investigate defence methods for few-shot classifiers. Under scenario (ii), where the support set of the target class is also randomised, we see lower transfer ASR across the chosen classes. We would like to remind readers that the adversarial samples were optimised explicitly using setting (i) and not for (ii). Even though the ASR in scenario (ii) is lower than in (i), there still exist classes, where unpleasantly high ASR occurs. 

\subsection{Detection of Adversarial Supports}

We compared our explored approaches against a simple filtering function for $r(\cdot)$, since prior detection methods for adversarial samples in few-shot classifiers do not exist. We experimented with using normal distributed noise as a filter, in which we computed the channel-wise variance for drawing normal distributed noise to be added to the images.
Being in the context of detection, we report the AUROC scores to evaluate the effectiveness of our detection algorithm.

\begin{figure}[htb]
    \centering
    \begin{subfigure}{0.47\textwidth}
        \includegraphics[trim={8.cm 53.2cm 9cm 5.5cm},clip,width=\textwidth]{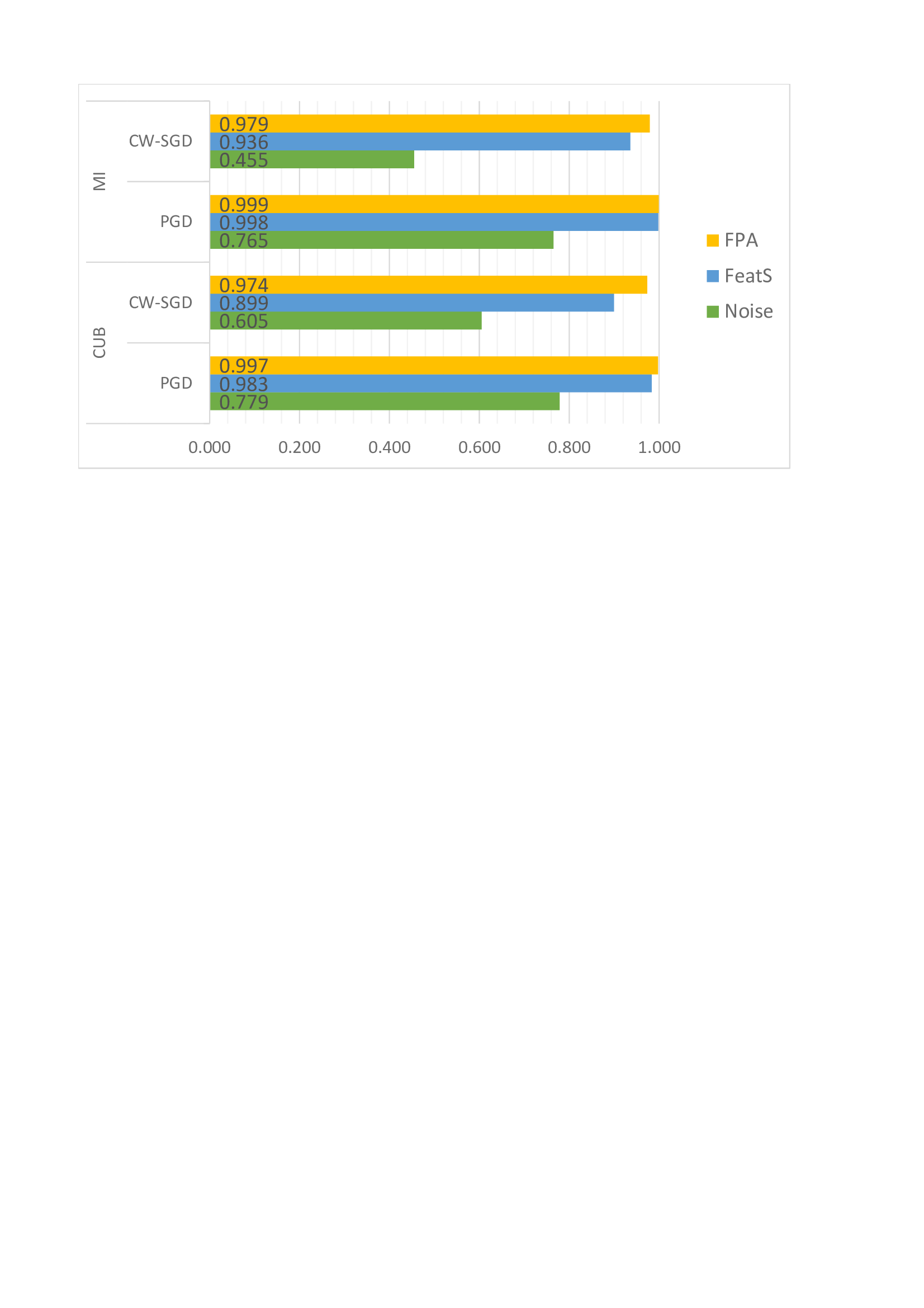}
        \caption{RN (5-shot) model.}
        \label{fig:rn}
    \end{subfigure}
    ~
    \begin{subfigure}{0.47\textwidth}
        \includegraphics[trim={8.cm 53.2cm 9cm 5.5cm},clip,width=\textwidth]{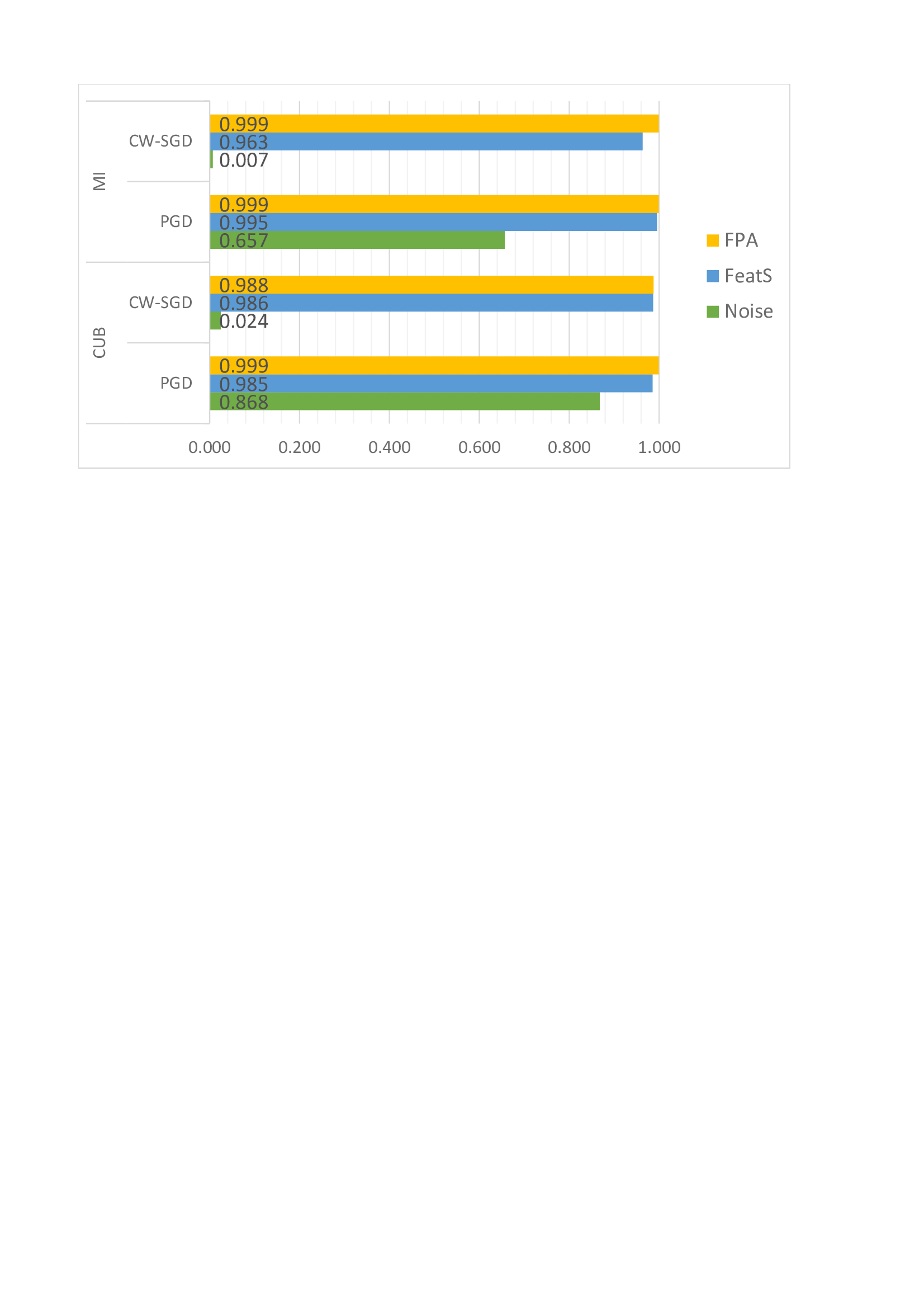}
        \caption{CAN (5-shot) model.}
        \label{fig:rn}
    \end{subfigure}
    \caption{Area Under the Receiver Operator Characteristic (AUROC) scores for the various filter functions (normal distributed noise, median filtering from Feature Squeezing (FeatS), FPA) across our experiment settings, for RN and CAN models. Higher is better.}
    \label{fig:aucdetection}
\end{figure}

Our results in Figure~\ref{fig:aucdetection} shows that FPA exhibits good detection performance.
\changemarker{This indicates that the self-similarity of clean samples under filtering of the auxiliary support set is preserved to a degree, which allows discrimination against adversarial samples.}
Though "FeatS" exhibits already a good detection performance, our FPA approach consistently outperforms it across all settings. 
The "Noise" approach, however, does not detect well. We see highly varied detection performances across the different settings, which makes this approach highly unreliable\footnote{Cases with AUROC score less than 0.5 indicates that more favourable detection effectiveness can be achieved by flipping the detection threshold (i.e. $U_{adv} > T$ to $U_{adv} < T$). However, it will not be experimentally consistent. This is also a clear indication of the lack of reliability of using "Noise" as a filtering function.}. This result is hardly surprising since such methods require substantial manual fine-tuning of its noise parameters. This is not ideal as newer attacks can be introduced in the future and also, being in a few-shot framework, the optimal noise parameters between different task instances might not be consistent as the data might be different. However, our FPA filter approach exhibits such robustness even in such scenarios as it still achieved favourable AUROC scores. For clean samples, our FPA managed to reconstruct $S_{aux}^c$ such that the logits of $Q_{aux}^c$ before and after filtering remained consistent, even when the FPA did not encounter classes from the novel split during training. 





\section{Discussion}
\label{discuss}

\subsection{Study of Self-Similarity Computation Methods}



\begin{figure}[htb]
    \centering
    \includegraphics[trim={48.3cm 101.5cm 4.5cm 4.5cm},clip,width = 0.35\textwidth]{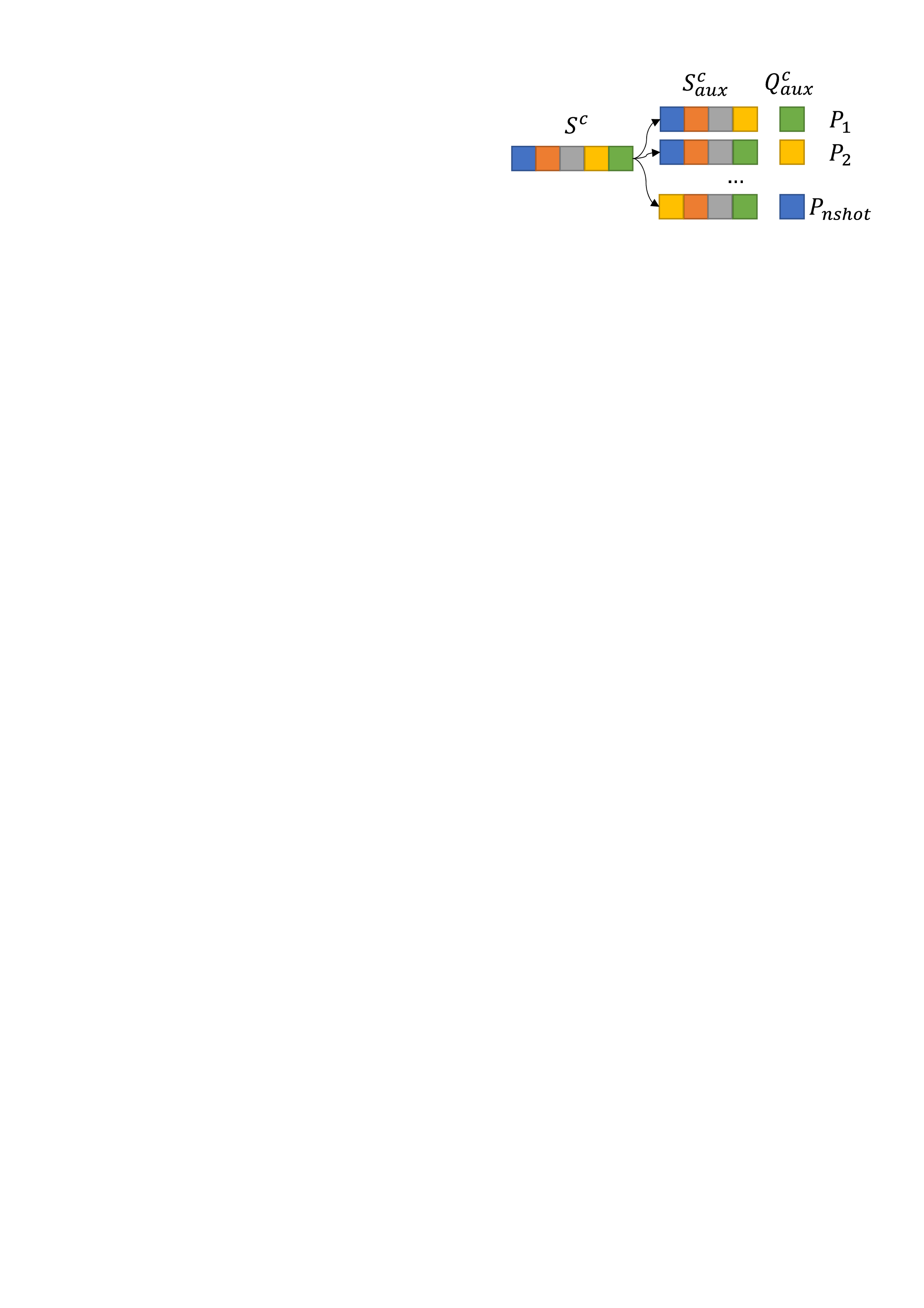}
    \caption{Illustration of how partitioning $S^c$ into two auxiliary sets $S^{c}_{aux}$ and $Q^{c}_{aux}$ is performed. Best viewed in colour.}
    \label{fig:partition}
\end{figure}

In Section~\ref{detectsec}, we described one of the possible detection mechanisms based on logits differences. An alternative would be to use hard label predictions. Thus, we investigate the effect of a differing scheme as a justification for our choice $U_{adv}$. For the case of hard label predictions, we perform the following: 
we compute the average accuracy of $Q^{c}_{aux}$, across the different permutated partitions of $S^{c}$, illustrated in Figure~\ref{fig:partition}. 
This results in the statistic $U_{adv}'$:
\begin{equation}
U_{adv}' = \frac{1}{n_{shot}} \sum^{n_{shot}}_{i=1} \mathbbm{1} [argmax(h(r(S^{c}_{i, aux}), Q^{c}_{i, aux})) \neq c] ,
\label{ssrate2}
\end{equation}
where $h$ is the few-shot classifier, $r$ is the filtering function, and $\mathbbm{1}$ is the indicator function.
Similarly, we flag the support set as adversarial when $U_{adv}'>T$, such that it goes beyond a certain threshold. 

\begin{table}[htb]
\centering
\resizebox{0.8\columnwidth}{!}{%
\begin{tabular}{|c|c|cc|cc|}
\hline
\multirow{2}{*}{Model}                                                  & \multirow{2}{*}{Dataset} & \multicolumn{2}{c|}{PGD} & \multicolumn{2}{c|}{CW-SGD} \\ \cline{3-6} 
                                                                        &                          & $U_{adv}$  & $U_{adv}'$    & $U_{adv}$  & $U_{adv}'$     \\ \hline
\multirow{2}{*}{\begin{tabular}[c]{@{}c@{}}RN\\ (5-shot)\end{tabular}}  & MI                       & \textbf{0.999}	&	0.451		&		\textbf{0.979} 	&	0.723  \\
                                                                        & CUB                      & 	\textbf{0.997}	&	0.326	&	\textbf{0.974}  		& 0.524	 \\ \hline
\multirow{2}{*}{\begin{tabular}[c]{@{}c@{}}CAN\\ (5-shot)\end{tabular}} & MI                       & \textbf{0.999}	&	0.991	&	\textbf{0.999}	&   0.931	   \\
                                                                        & CUB                      & \textbf{0.999}		&	0.998	&	\textbf{0.988}	&	0.821  \\ \hline
\end{tabular}
}
\caption{Area Under the Receiver Operator Characteristic (AUROC) scores for the two detection mechanisms ($U_{adv}$ and $U_{adv}'$) using our FPA across our experiment settings. Higher is better.}
\label{tab:aucdetectionDM}
\end{table}

Table~\ref{tab:aucdetectionDM} shows our AUROC scores comparing the two detection mechanisms, $U_{adv}$ and $U_{adv}'$, when using the FPA filtering function. It is evident that using logits scores to calculate differences as in $U_{adv}$ can be more informative than using hard label predictions to match class labels, as $U_{adv}$ consistently outperforms $U_{adv}'$, with the difference being bigger for RN. Differences in logits can be pronounced also in cases when the prediction label does not switch. 

\subsection{Varying Degrees of Regularisation of FPA}
We observe lower AUROC scores for the RN model than the CAN model in Figure~\ref{fig:aucdetection}. As such, we question if this difference can be attributed to the FPA's ability to reconstruct clean samples effectively.
Recalling from Eq.~\eqref{aeeqn}, we define an additional regularisation term to enforce stricter reconstruction requirements to also include class distribution reconstruction. More specifically, we minimise the following objective function: 

\begin{equation}
\begin{split}
\mathcal{L_{FPA'}} =\frac{1}{N'}\sum^{N'}_{i=1} 0.01 &\cdot \ \frac{\| x_i - \hat{x_i} \|_{2}^{2}}{dim(x_i)^{1/2}} \ + \frac{\| f_i - \hat{f_i} \|_{2}^{2}}{dim(f_i)^{1/2}}\\
&+ \frac{\| z_i - \hat{z_i} \|_{2}^{2}}{dim(z_i)^{1/2}} \ ,
\end{split}
\label{aeeqn2}
\end{equation}
where $x_i$ and $\hat{x_i}$ are the original and reconstructed image samples, respectively, and $f_i$ and $\hat{f_i}$ are the feature representation of the original and reconstructed image obtained from the few-shot model before any metric module, and $z_i$ and $\hat{z_i}$ are the logits of the original and reconstructed image. We refer to this variant as $FPA'$.
Similarly, we train $FPA'$ by fine-tuning the weights from the standard autoencoder.

\begin{table}[htb]
\centering
\resizebox{0.6\columnwidth}{!}{%
\begin{tabular}{|c|cc|cc|}
\hline
\multirow{2}{*}{Dataset} & \multicolumn{2}{c|}{PGD}  & \multicolumn{2}{c|}{CW-SGD} \\ \cline{2-5} 
                         & $FPA$ & $FPA'$ & $FPA$ & $FPA'$  \\ \hline
MI             &     0.999    &   0.999    &     0.979    &   0.950     \\ \hline
CUB                      &   0.997      &  0.997     &    0.974      &  0.971       \\ \hline
\end{tabular}
}
\caption{AUROC results comparing $FPA$ and $FPA'$ for the RN. We computed the results across the 10 and 25 exemplary classes for MI and CUB respectively, and 50 sets of adversarial perturbations. 
}
\label{tab:scratedirtycompare}
\end{table}

Our results in Table~\ref{tab:scratedirtycompare} shows that surprisingly, imposing a higher degree of regularisation marginally lowers the detection performance of our algorithm rather than improving it. This implies that $FPA$ is already sufficient to induce a large enough divergence in classification behaviours in the presence of an adversarial support set.

\section{Conclusion}
\label{conclude}
Adversarial attacks against the support sets can be damaging to few-shot classifiers. 
To this end, we propose a novel adversarial attack detection algorithm on support sets in the few-shot framework, which has not been explored prior, to the best of our knowledge. Our algorithm works by using the concept of self-similarity among samples in the support set and filtering. We obtained high detection AUROC scores in the CAN and RN models, across MI and CUB datasets, with FPA and FeatS filtering functions, though FPA is superior. We have also found that using differences of the logits scores yield better detection performances and a higher degree of regularisation of FPA does not guarantee better detection results. 
Future work can explore the efficacy of our detection for black-box attack settings and the detection performances with different filtering functions.

\section{Acknowledgements}
This research is supported by both ST Engineering Electronics and National Research Foundation, Singapore, under its Corporate Laboratory @ University Scheme (Programme Title: STEE Infosec-SUTD Corporate Laboratory).

\bibliographystyle{named}
\bibliography{ijcai21}

\end{document}